\documentclass[reprint, amsmath,amssymb,showpacs, aps,prd]{revtex4-1}
\usepackage[utf8]{inputenc}
\usepackage{amsmath}
\usepackage{tikz}
\usepackage{amsfonts}
\usepackage{amssymb}
\usepackage{mathrsfs}
\usepackage{latexsym}
\usepackage{footnote}
\usepackage{hyperref}
\usepackage[english]{babel}
\usepackage{cmap}
\usepackage[T1]{fontenc}
\usepackage{textcomp}
\usepackage{setspace}
\usepackage{graphicx}
\usepackage{dcolumn}
\usepackage{bm}
\usepackage{float}
\begin{document}

\preprint{APS/123-QED}

\title{RESONANT RADIOFREQUENCY FIELDS DAMAGING \textit{SACCHAROMYCES CEREVISIAE}}

\author{$^1$W S Dias}
\author{$^1$E H M Liquer}
\author{$^2$L C Gontijo}
\author{$^2$T A Oakes}
\author{$^2$G S Dias}
\author{$^2$C Marques}
\author{$^3$H S Chavez}
\affiliation{$^1$WS-Tecnologia,  Serra-ES, Brazil}
\affiliation{$^2$Department of Physics - IFES - Instituto Federal do Esp\'{i}rito Santo (IFES), \\ Av. Vit\'{o}ria 1729, Jucutuquara, Vit\'{o}ria, ES, Brazil, CEP 29040-780} 
\affiliation{$^3$UCL-Faculdade do Centro Leste, Rodovia ES 010, Km 6 \\ BR 101, Serra-ES, Brazil}

\email{celiom@ifes.edu.br}
\email{gilmar@ifes.edu.br}

\date{\today}

\begin{abstract}
This work describes an experiment and some results from instability involving the  proliferation achieved by cells over some radio frequency electromagnetic field. Saccharomyces cerevisiae cells were cultured, and separate samples were examined within $6$ hours. The frequency, the pulse width and peak-to-peak voltages were fixed. Producing at least $1$ kV across the cell membrane in milk suspension. It was observed that in the presence of the electromagnetic field the cells had their time life  drastically reduced when compared to the control sample.
\end{abstract}


\maketitle


\section{\label{sec:level1}Introduction}

 The use of electric fields to control the proliferation of microorganisms were first described and patented by \cite{art1} in the early 1960s. After, reference \cite{art2}  analyzed the effects of DC pulses on microorganisms in a systematically way. Following these works, radio frequency technology has proven useful for a wide range of applications over the decades since the initial research. The extensive use of electromagnetic field has been applied in food preservation as discussed in \cite{art3,art4,art5,art6,art7}.
    The authors describe industrial processes with the application of radio frequencies for the elimination of microorganisms in food production, in order to increase shelf life and final product quality.

Other applications are: the interruption of the formation processes or even the synchrony of the cellular reproduction process, as it is well known. According to them, only by varying the potential of the field applied, one can stimulate and inhibit cellular processes.

The main goal of this work is to study the effects of the exposure 
of high frequency electromagnetic waves in a solution of milk and Saccharomyces cerevisiae and to evaluate the effects of fields.
\section{Presentation of the Experiment and main Results}

Saccharomyces Cerevisiae S$288$C is an unicellular eukaryotic organism that belongs to the kingdom of fungi. It is the yeast used in the production of bread and also of beer, besides being used for the production of ethanol. A milk solution containing $20$ g of dry biological yeast (Saccharomyces cerevisiae), $20$ g of sucrose and $200$ ml of milk was prepared at $28$ Celsius degrees. The solution was divided into four equal samples, two for control and two for the application of the electromagnetic field.
We use the Radio Frequency Generator Kenwood Transceiver, Model TS-$2000$, operating in a constant frequency mode of $28$ MHz frequency carrier modulation. This frequency was chosen because it was part of the used range in the cited studies involving radio frequency.
The antenna that was developed for our purposes  is called magnetic loop and is shown at figure $1$. It constitutes of an RLC circuit with a metal circle in a torus geometry, which resonates at the frequency of the experiment. Measurements were calculated in the MMANA antenna simulator software, obtained at \cite{art10}, following the field setup in order to expose the samples. See figure below.
\begin{figure}[h!]
\centering
\includegraphics[scale=0.32]{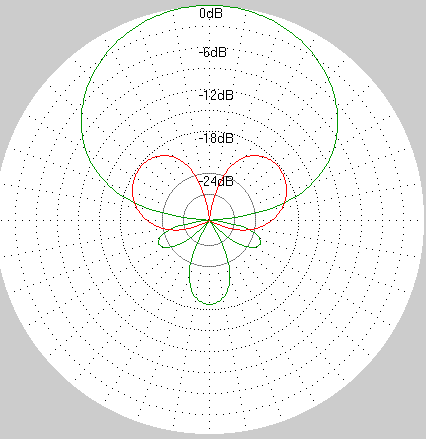}
\caption{ Field geometric setup of an antenna.}
\label{fig:stream}
\end{figure}
 
Based on these materials, the antenna was placed together with the radio frequency generating equipment on a wooden table covered by glass. Inside the loop of the antenna - area of the greatest electromagnetic field - two beakers were arranged with $50$ ml of the solution in each one. Two additional $50$ ml beakers were placed outside and away from the antenna loop, spaced at a distance of at least one meter from the radiating assembly and adopted as a control sample.

A power of 25 watts was applied to the antenna, bringing the peak-to-peak voltage at the coil to 2.8 v. In the figure below it can be verified that a fluorescent lamp remains lit without any electrical contact inside the loop, demonstrating the high intensity of the lines of force that concentrate there. The experiment environment was closely supervised to verify that the temperature remained the same as the initial temperature. In the temperature measurements performed every hour, during the 6 hours of the experiment, no difference in temperature between the samples was noticed. This precaution was taken so that the temperature variation did not interfere in the development of the cells in the analysis of the result.
After 6 hours of exposure, individual samples of each beaker were collected and survivor cells were counted throughout photos captured from a microscope. The general state of the cells of each beaker could be visualized as shown at the figures$2$ and $3$ in the next section.
\begin{figure}[h!]
\centering
\includegraphics[scale=0.5]{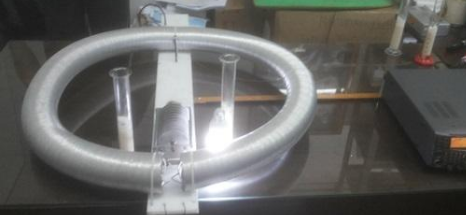}
\caption{Experiment setup showing the control and treated samples.}
\label{fig:stream}
\end{figure}

Principal results are shown at the figures below, where we see a stringent correlation between electromagnetic field application and mortality of the micro organisms.   

\begin{figure}[h!]
\centering
\includegraphics[scale=0.3]{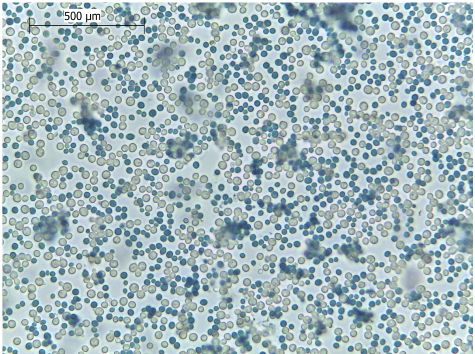}
\caption{Microscopic image of the cell sample after field exposure: the majority of the cells are partially inactive (light blue ones) or completely inactive (dark blue ones).}
\end{figure}

The control sample is showed at figure 3 below.

\begin{figure}[h!]
\centering
\includegraphics[scale=0.3]{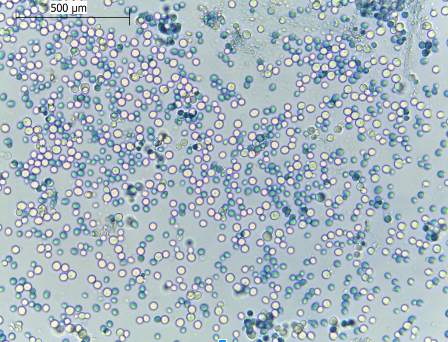}
\caption{Microscopic image of the control cell sample: the majority of the cells are alive (white ones) or partially inactive (light blue ones).}
\end{figure}

At the bar graph (figure 4), we summarize the results. The first blue colored bar, represents the percentage number of live microorganisms (L)  at the control sample, the second one (green) represents the percentage of dead-live (DL) cells and the third (yellow) represents the percentage of dead cells. At the second set of colored blue, green and yellow bars, we see a completely distinct behavior in the sample which was submitted to the electromagnetic field. Clearly we have a growth of the population of the affected cells instead of the live ones. 
\begin{figure}
\centering
\includegraphics[scale=0.4]{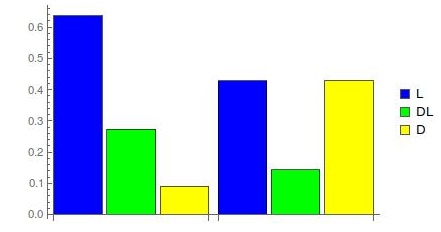}
\caption{The bars indicate the percentage of the survivor cells (blue bars), the sick cells (green bars) and the dead cells (yellow bars). The first three  bars are the control sample, the remaining are after field exposure.}
\end{figure}

It is possible to interpret these results in terms of an elegant simple model, which describes these initial results in a zero order approximation, not in quantum level approach, by coupling  the external field with the ion charge density inside the cell. 

In fact, it is possible to define a classical resonant flux model when we consider the interaction of the external electromagnetic field (EEF) with the ions flux inside the cell (at cell membrane or at the mitochondria) responsible for production of ATP. The resonances could happen when external electromagnetic field frequency $w_{EEF}$  is an integer multiple of the ion flux frequency $w_{ions}$. 

In the resonance case the intensity of the ions flux increases, producing an elevation of the cell ATP. When $w_{EEF}$ is different of an integer multiple  of the $w_{ions}$, the intensity of the ions flux of the cell tends to be annihilated, thus decreasing the ATP. In both cases, of the resonance, or not, the ATP will change; the intensity of this change depends on the Poynting vector of the EEF.

\section{Conclusions}
As predicted by \cite{art8,art9} the immersion in the electromagnetic field proved to be efficient in drastically reducing the number of healthy cells in the samples that were submitted to the field. Such thing could be happening through interference in the mitochondrial electrochemical processes or by the disruption of the cell's membrane integrity.

We devise, following \cite{art8} that the long exposition of electromagnetic field may cause a variation of the ATP demand for the basic process inside the cells, stressing them and killing cells. The basic physics behind this process can be related to a dynamically redefinition of ion concentrations or ion gradients inside the cells. This hypothesis needs deeper analyses and future experiments. Which are being considered for future developments. 

As we see, a wide range of applications not only in the area of food preservation but also in the medical field can be glimpsed from this technique, enabling the development of industrial and medical equipment aimed at microbiological control. 
Other possible application is to try to reduce the tumoral cell growth by applying specific fields setups that, in some way, could fully break mitochondrial metabolism.  

In particular, we are developing the study of the effect of radiation at the proliferation of more complex organisms, whose large reproduction causes diseases at agricultural plants and forests like the Atlantic Forest in Espírito Santo, Brazil.

\begin{acknowledgments}
We thank Dr. Adriana Korres from Federal Institute, IFES, for the support in the experiment at the Microbiology Lab. Also we thank Isabelle Cohen Dias for the suggestions in the writting of this paper. 
\end{acknowledgments}

\end{document}